\documentclass[aip,reprint,twocolumn,showpacs,superscriptaddress]{revtex4-1}  
\usepackage{units}
\usepackage{amssymb}
\usepackage{graphicx}
\usepackage{esint}
\usepackage{xcolor}

\newcommand{\pvec}[1]{\vec{#1}\mkern2mu\vphantom{#1}}

\begin{document}










\title{Universal features of phonon transport in nanowires with correlated surface roughness}

\author{L. N. Maurer}
\email{lnmaurer@wisc.edu}
\affiliation{University of Wisconsin-Madison, Madison, Wisconsin 53706, USA}

\author{Z. Aksamija}
\affiliation{University of Massachusetts Amherst, Amherst, Massachusetts 01003, USA}

\author{E. B. Ramayya}
\altaffiliation{Currently with Intel Corporation, Hillsboro, OR, USA}
\affiliation{University of Wisconsin-Madison, Madison, Wisconsin 53706, USA}

\author{A. H. Davoody}
\affiliation{University of Wisconsin-Madison, Madison, Wisconsin 53706, USA}

\author{I. Knezevic}
\email{knezevic@engr.wisc.edu}
\affiliation{University of Wisconsin-Madison, Madison, Wisconsin 53706, USA}

\begin{abstract}

The ultralow thermal conductivity $\kappa$ observed experimentally in intentionally roughened silicon nanowires (SiNWs) is reproduced in phonon Monte Carlo simulations with exponentially correlated real-space rough surfaces similar to measurement [J. Lim, K. Hippalgaonkar, S. C. Andrews, A. Majumdar, and P. Yang, Nano Lett. 12, 2475 (2012)]. Universal features of thermal transport are revealed by presenting $\kappa$ as a function of the normalized geometric mean free path $\bar\lambda$ ($0<\bar\lambda<1$); the diffusive (Casimir) limit corresponds to $\bar\lambda=1/2$. $\kappa$ vs $\bar\lambda$ is exponential at low-to-moderate roughness (high $\bar\lambda$), where internal scattering randomly interrupts phonon bouncing across the SiNW, and linear at high roughness (low $\bar\lambda$), where multiple scattering events at the same surface results in ultralow, amorphous-limit thermal conductivity.
\end{abstract}

\maketitle
Heat flow through semiconductor nanostructures is governed by phonons, the quanta of lattice waves. \cite{Ziman_Book} Nanoscale phonon transport is presently a very active field of research, \cite{CahillJAP03,CahillAPR2012} straddling basic science inquiry  \cite{RoukesNature2000_QuantumThermalCond,GhoshNatMat2010,BaeNatComm2013} and applications in electronics, \cite{PopReviewNanoRes,SinhaJHT2005,RalevaIEEETED2008} optoelectronics,  \cite{VitielloIEEEJSTQE2008,YaoNatPhot2012,ShiJAP2014} and thermoelectrics. \cite{VineisAdvMat2010,PYang_Nat_08,MajumdarScienceReview} Aided by considerable advances in measurement,  \cite{WeathersShi2013Measurement,ChenACSNano2011} the design space for thermal transport has been greatly expanded by the use of low-dimensional and nanostructured semiconductors, \cite{BalandinMT,PhononicsRMP2012} which range from quasiballistic graphene-based systems with superior heat conduction \cite{BalandinGrapheneReviewNatMat2011,ChenGrapheneNatMat2012,BaeNatComm2013} to structures in which rough surfaces or interfaces lead to low and anisotropic thermal conductivity, \cite{GargChen2013,AksamijaPRB2013,AksamijaAPL2011} such as superlattices, \cite{SimkinMahanPRL,ChenPRB98,GargChen2013,AksamijaPRB2013} nanomembranes,  \cite{BalandinWangPRB98_ThinFilm,Majumdar_JHT_01,Goodson_JAP_02,Aksamija_PRB_10,KaramitaheriJAP2013}, nanowires (NWs), \cite{Majumdar_APL_03,PYang_Nat_08,Shi_APL_08,PYang_NLet_12,FeserCahillJAP2012_NanowireArrays} and structures with nanodots.\cite{Pernott_NatureMat10_QD,Nika_PRB11_QDSuperlatt}

However, theoretical understanding of phonon dynamics in nanostructures with a significant degree of disorder is far from complete. A prominent open problem is the unexpectedly low thermal conductivity, $\kappa$, of very rough silicon nanowires (SiNW). \cite{PYang_Nat_08,PYang_NLet_12,FeserCahillJAP2012_NanowireArrays} Earlier measurements on vapor-liquid-solid (VLS)-grown NWs showed $\kappa$ an order of magnitude lower than the bulk value. \cite{Majumdar_APL_03} These results could largely be explained within the diffusive-transport framework and a simple model of phonon interaction with surfaces: \cite{Casimir_38, Ziman_Book} the rough surface is described by a specularity parameter $p$, which is the probability that a phonon would reflect specularly upon impact. The diffusive or Casimir limit corresponds to complete momentum randomization at the surface ($p=0$). Indeed, thermal transport in many relatively smooth NWs is well described using the relaxation time approximation and the model of partially or completely diffuse surface scattering.\cite{Mingo_PRB_03,Bera_JAP_12,Majumdar_APL_03,Majumdar_JHT_05,CahillAPR2012} In contrast, the $\kappa$'s measured on electrolessly etched \cite{PYang_Nat_08} or intentionally roughened VLS-grown \cite{PYang_NLet_12} NWs are far below the Casimir limit and have values similar to amorphous materials. Several groups have calculated the $\kappa$ of SiNWs in the presence of periodic or rough surface features by molecular dynamics,  \cite{CCYang_CMS_11, Sansoz_NLet_11, Galli_PRL_12} elastodynamics or hydrodynamics, \cite{Cross_PRL_2001, Gong_PRB_09, Jang_JAP_12,Sellitto_JHT_11} solving the phonon Boltzmann transport equation (BTE), \cite{Majumdar_JHT_05, Shi_APL_08, YChen_PhysicaB_11,
Bera_JAP_12, Bourgeois_APL_13,Bourgeois_NLet_09, Schmidt_PRB_12,Ramayya_PRB_12} or within coherent transport formalisms. \cite{Mingo_PRB_11,Moore_PRB_07, Majumdar_PRL_08, Sinha_PRB_11,Sinha_NLett_13} The unusually low thermal conductivity has been linked to diverse mechanisms, such as the need to include full dispersion, \cite{Majumdar_JHT_05} different surface-scattering rates for phonons of different energies, \cite{Majumdar_PRL_08,Ravaioli_PRL_09} a surrounding native oxide that might  randomize phonon phase,  \cite{DonadioPRL2009,CahillAPR2012} internal defects,  \cite{FeserCahillJAP2012_NanowireArrays,Galli_PRL_12} and multiple coherent backscattering events from highly correlated surfaces.  \cite{Sinha_PRB_11,Sinha_NLett_13} 
Recently, Lim \textit{et al.} \cite{PYang_NLet_12} performed a systematic experimental study of the surface condition of intentionally roughened  VLS-grown NWs, revealing correlation lengths smaller than previously assumed \cite{Sinha_PRB_11} and underscoring the importance of surface-roughness profile in nanoscale thermal transport.

In this letter, we show that the ultralow measured thermal conductivities in intentionally roughened SiNWs can be reproduced in phonon Monte Carlo simulation with exponentially correlated real-space rough surfaces similar to experiment. \cite{PYang_NLet_12} We introduce the \textit{normalized geometric mean free path} $\bar\lambda$, a dimensionless quantity proportional to the NW volume-to-surface ratio, which serves as a universal quantifier of surface roughness; $\bar\lambda =1/2$  corresponds to the Casimir limit. $\kappa(\bar\lambda)$  is exponential at low roughness (high $\bar\lambda$), where internal scattering competes with roughness scattering, and linear at pronounced roughness (low $\bar\lambda$), where multiple successive scattering events from the same surface dominate and result in ultralow, amorphous-limit $\kappa$.

\begin{figure}
\includegraphics[width=0.8\columnwidth]{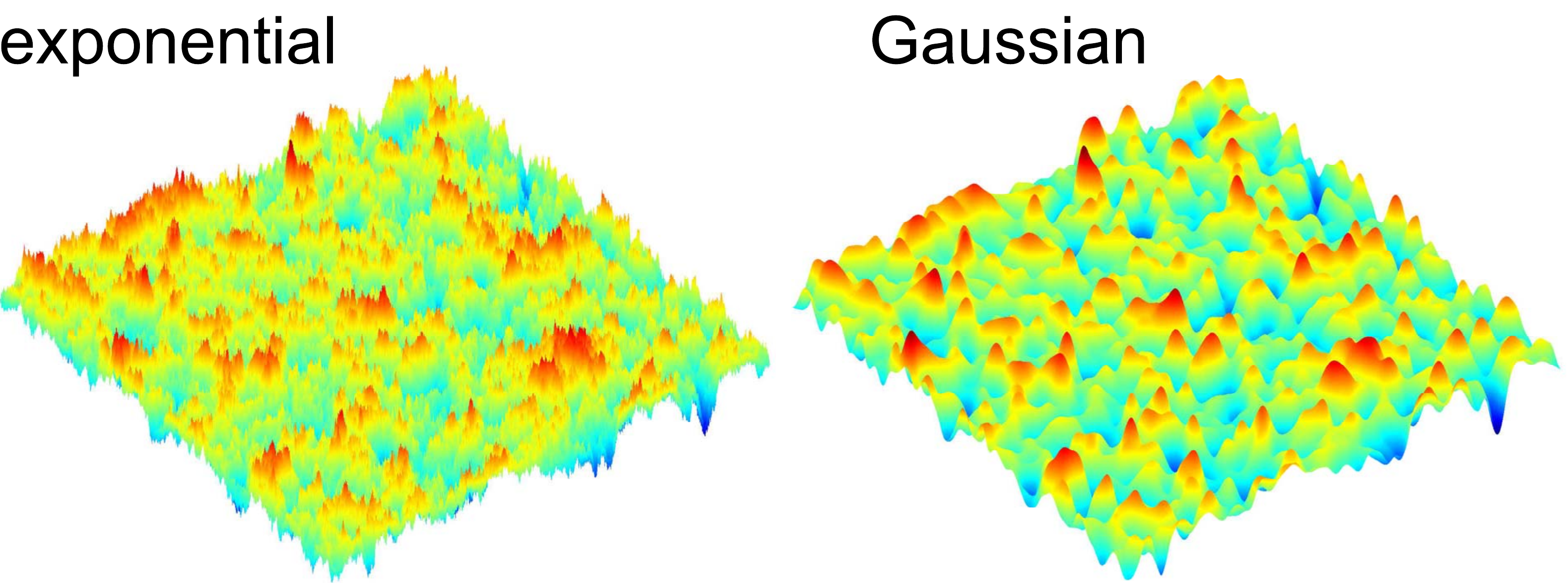}
\caption{Example surface profiles with exponential (left) and Gaussian (right) autocorrelation functions. They have the same rms roughness (2 nm) and
correlation length (25 nm, chosen large for visual appeal), and were created from the same set of random numbers.}
\label{fig:Example-Surfaces}
\end{figure}

We consider long SiNWs with square cross-sections and rough surfaces characterized a given autocorrelation function (ACF). If $h\left(\vec{r}\right)$ is the height of surface $S$ at point $\vec{r}$, then
the ACF is given by $C\left(\vec{r}\right) = \iint_S h\left(\pvec{r}'\right)
h\left(\pvec{r}' + \vec{r}\right) d^2\pvec{r}'$. Most previous studies that included correlated roughness used either
Gaussian \cite{Cross_PRL_2001,Ravaioli_PRL_09, Sinha_PRB_11, Jang_JAP_12,
Lee_JAP_12} or Gaussian-based \cite{Gong_PRB_09} ACFs. However, measurements of Si/SiO$_2$ interfaces found exponential \cite{Goodnick_PRB_85} or
exponential/Gaussian hybrid ACFs. \cite{Takagi_TED_10} Hybrid ACFs have also been found on SiNWs. \cite{Sinha_NLett_13, PYang_NLet_12} Therefore, we consider surfaces with both exponential [$C_e\left(\vec{r}\right) =
\Delta^2 e^{-\left|\vec{r}\right|/\xi}$] and Gaussian [$C_g\left(\vec{r}\right) = \Delta^2 e^{-\left|\vec{r}\right|^2/\xi^2}$] ACFs (exponential and Gaussian surfaces, for brevity), where $\Delta$ denotes the rms roughness and $\xi$ is the correlation length. In the simulation, each NW surface has a random roughness profile generated numerically according to a given ACF, $\Delta$, and $\xi$.~\cite{Buran_IEEE_09, Ramayya_PRB_12}  Figure \ref{fig:Example-Surfaces} illustrates the differences between exponential and Gaussian surfaces that have the same $\Delta$ and $\xi$. The two surfaces have similar large-scale features, but the exponential surface has considerably more small-scale roughness.

The SiNWs we consider have widths 20-70 nm and are approximately 2 $\mathrm{\mu m}$ long, much longer than the room-temperature bulk phonon mean free path, \cite{JuGoodsonAPL99} so transport is scattering-limited (diffusive) and described by the phonon BTE. \cite{Ziman_Book, Carruthers_RMP_61} Phonon Monte Carlo (PMC) is an efficient stochastic technique for the solution to the BTE,  \cite{Peterson_JHT_94, Majumdar_JHT_01, Lacroix_PRB_05,
Baillis_JHT_08, Ramayya_PRB_12} which can incorporate real-space roughness and simulate the SiNWs of experimentally relevant sizes. In PMC, a large ensemble of numerical phonons is tracked over time as they fly freely and undergo scattering from the real-space surface roughness (upon hitting a point on the rough surface, the phonon reflects specularly) or from internal scattering mechanisms (normal, umklapp, and isotope scattering; we used the rates from Morelli \textit{et al}. \cite{Slack_PRB_02}). The simulation includes longitudinal and transverse acoustic phonons; optical phonons carry little heat owing to their low group velocity, but they affect the scattering rates of acoustic phonons and this influence is included in the above rates. Bulk dispersions are reasonable for NWs thicker than a few tens of nanometers \cite{Mingo_PRB_03,Majumdar_JHT_05,Bera_JAP_12} and, for simplicity, we use isotropic relationships for each acoustic branch. \cite{Pop_JAP_04} Planar black-body contacts, whose temperatures differ by 20 K, are at the two ends of the wire. \cite{Baillis_JHT_08} With the contact temperatures given, the simulation runs until a steady state is reached, as evidenced by a linear temperature profile (constant temperature gradient $\nabla T$) and a constant power flux $\vec\Phi$ along the NW. $\kappa$ is obtained from Fourier's law, $\vec\Phi=-\kappa \nabla T$. Implementation details can be found in Refs. \onlinecite{Ramayya_PRB_12} and \onlinecite{Knezevic_JAP_14}.



\begin{figure}
\includegraphics[width=\columnwidth]{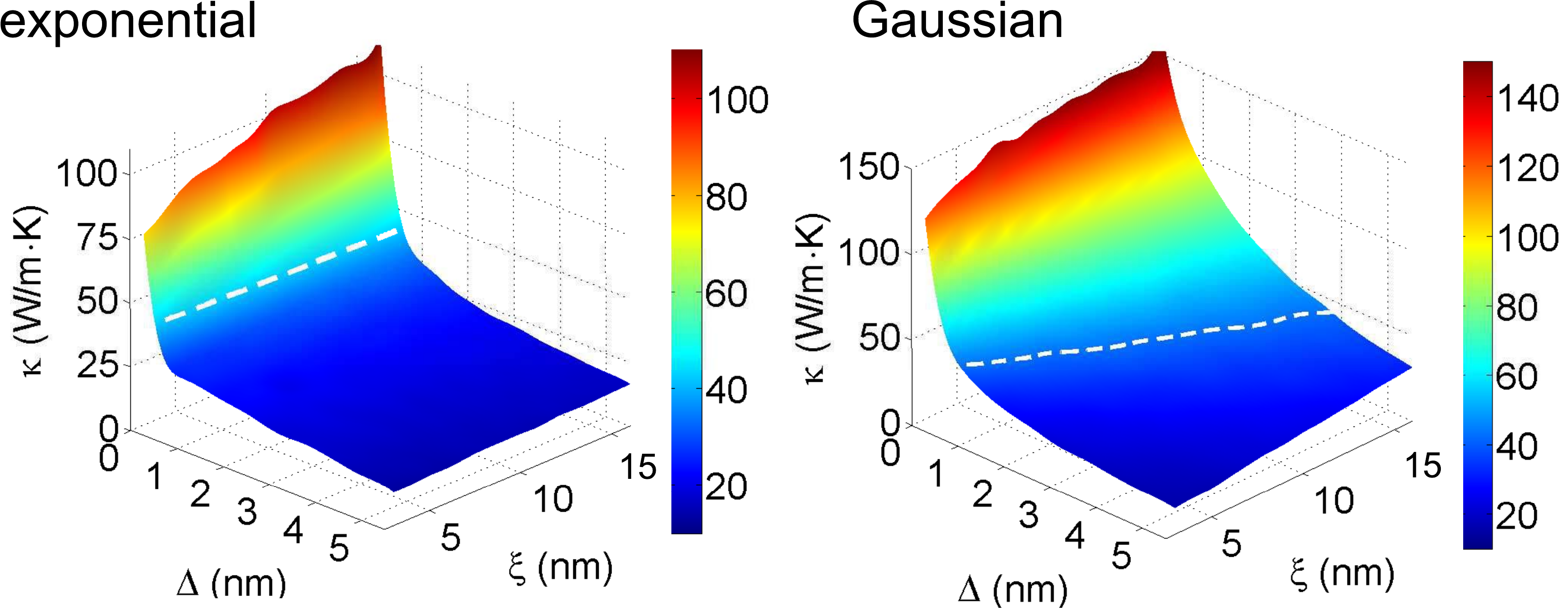}
\caption{Room-temperature thermal conductivity $\kappa$ versus rms roughness $\Delta$ and correlation length $\xi$ for 70-nm-wide SiNWs with exponential (left panel) and Gaussian (right panel) autocorrelation functions. The Casimir limit for SiNWs of this width, approximately 42.7 W/m$\cdot$K, is shown with a dashed white line.}
\label{fig:Kappa-vs-rms-and-clen}
\end{figure}

Figure \ref{fig:Kappa-vs-rms-and-clen} presents the results of PMC simulations for a large ensemble of 70-nm-wide SiNWs, akin to those measured by Lim \textit{et al}. \cite{PYang_NLet_12} The rough surfaces were generated according to exponential or Gaussian ACFs, with a broad range of roughness parameters, $\Delta$=0-5.5 nm and $\xi$=3-16 nm; this range includes experimental values. \cite{PYang_NLet_12} As expected, $\kappa$ decreases with increasing $\Delta$ and increases with increasing $\xi$ (the surface ``looks" smoother with a higher $\xi$. \cite{Soffer_JAP_67}) Exponential surfaces have more small-scale roughness, and consequently lower $\kappa$'s, because they are more effective at scattering short-wave-length phonons than their Gaussian counterparts. Furthermore, $\kappa$ calculated with real-space roughness is below the Casimir limit (about 42.7 W/m$\cdot$K for 70-nm NWs and shown in a white dashed line in Fig. \ref{fig:Kappa-vs-rms-and-clen}) over a wide range of $\Delta$ and $\xi$ for both exponential and Gaussian surfaces. The Casimir limit was obtained from PMC, wherein the NW surfaces are flat but each phonon's momentum is completely randomized whenever it hits a surface. 

\begin{figure}
\includegraphics[width=0.8\columnwidth]{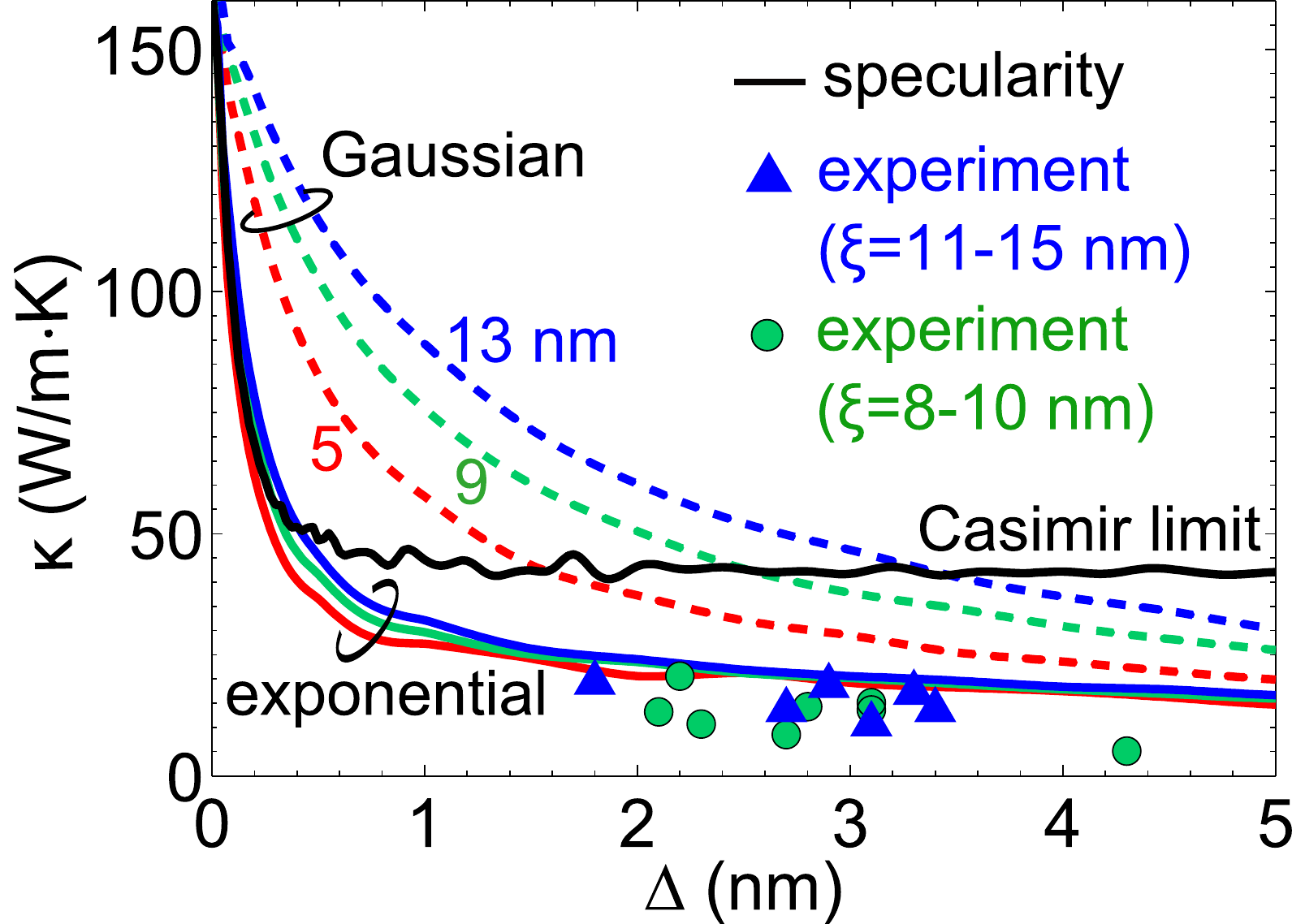}
\caption{Thermal conductivity $\kappa$ versus rms roughness $\Delta$ for 70-nm-wide SiNWs with different correlation lengths $\xi$ and autocorrelation functions of the surface roughness. Values from the PMC calculation with real-space exponential
(solid curves) and Gaussian (dashed curves) surfaces are presented for the correlation lengths of 5 nm (red), 9 nm (green), and 13 nm (blue). Thermal conductivity calculated by PMC with a momentum-dependent
specularity-parameter model for phonon surface-roughness
scattering (black solid curve) tends to the Casimir limit. The symbols represent experimental values from the supplement to Lim \textit{et al.}, \cite{PYang_NLet_12} obtained on intentionally roughened SiNWs of diameters 67-84 nm, and  with correlation lengths 8-10 nm (green circles) and 11-15 nm (blue triangles) extracted from exponential fits to experimentally obtained ACFs.} \label{fig:kappa-Soffer}
\end{figure}

Figure \ref{fig:kappa-Soffer} further illustrates the importance of accounting for real-space correlated roughness. The specularity parameter \cite{Ziman_Book, Soffer_JAP_67} $p$ is a wave concept that can be derived based on basic diffraction theory in the limit of no correlation as $p(\vec q)=\exp\left(-4\Delta^2 q_\perp^2\right)$, where $q_\perp$ is the component of the wave vector normal to the surface. \cite{Soffer_JAP_67,Aksamija_PRB_10} Figure \ref{fig:kappa-Soffer} shows $\kappa$ versus $\Delta$ for the specularity-parameter model ($\xi=0$) and for exponential and Gaussian surfaces with different values of $\xi$ (5, 9, and 13 nm). The exponential curves and the specularity model agree up to about $\Delta\approx 0.5\,\mathrm{nm}$, which is on the order of the average phonon wave length, after which the specular curve saturates at the Casimir limit ($p=0$). Moreover, our calculation results for exponential surfaces are very close to experiments on intentionally roughened NWs of similar diameters and surface roughness parameters. \cite{PYang_NLet_12} The slight discrepancy might stem from isotropic rather than full dispersions, \cite{Majumdar_PRL_08,Knezevic_JAP_14} the fact that the measured SiNWs appear to have a hybrid rather than purely exponential ACF, \cite{PYang_NLet_12} and not accounting for the native oxide. \cite{Galli_PRL_12}



Thermal conductivity includes the effects of both internal and surface scattering. Focusing on surface scattering alone, we note that a NW can be considered an open cavity. In a closed cavity, the geometric mean free path (GMFP), $\lambda$, is the average distance a phonon travels between successive surface scattering events in the absence of internal scattering and is computed simply as $\lambda = \frac{4V}{S}$, where $V$
is the cavity volume and $S$ is its surface area. \cite{Chernov_JSP_97} Our wires have only small openings at the ends, so they resemble closed cavities; numerical computation of $\lambda$ in our NWs shows them to be almost identical to the closed-cavity values. A long square NW of length $L$ has $V =L W^2$ (since the average height of rough surfaces is zero) and $S\geq 4 W L$ (the rougher the NW, the larger its surface area), thus $\lambda \leq W$. Therefore, we define the normalized geometric mean free path, $\overline{\lambda} = \lambda/W$ ($0\leq \bar\lambda\leq 1$), which enables us to compare across different wire diameters. Smaller $\bar\lambda$ means greater roughness.

\begin{figure}
\includegraphics[width=\columnwidth] {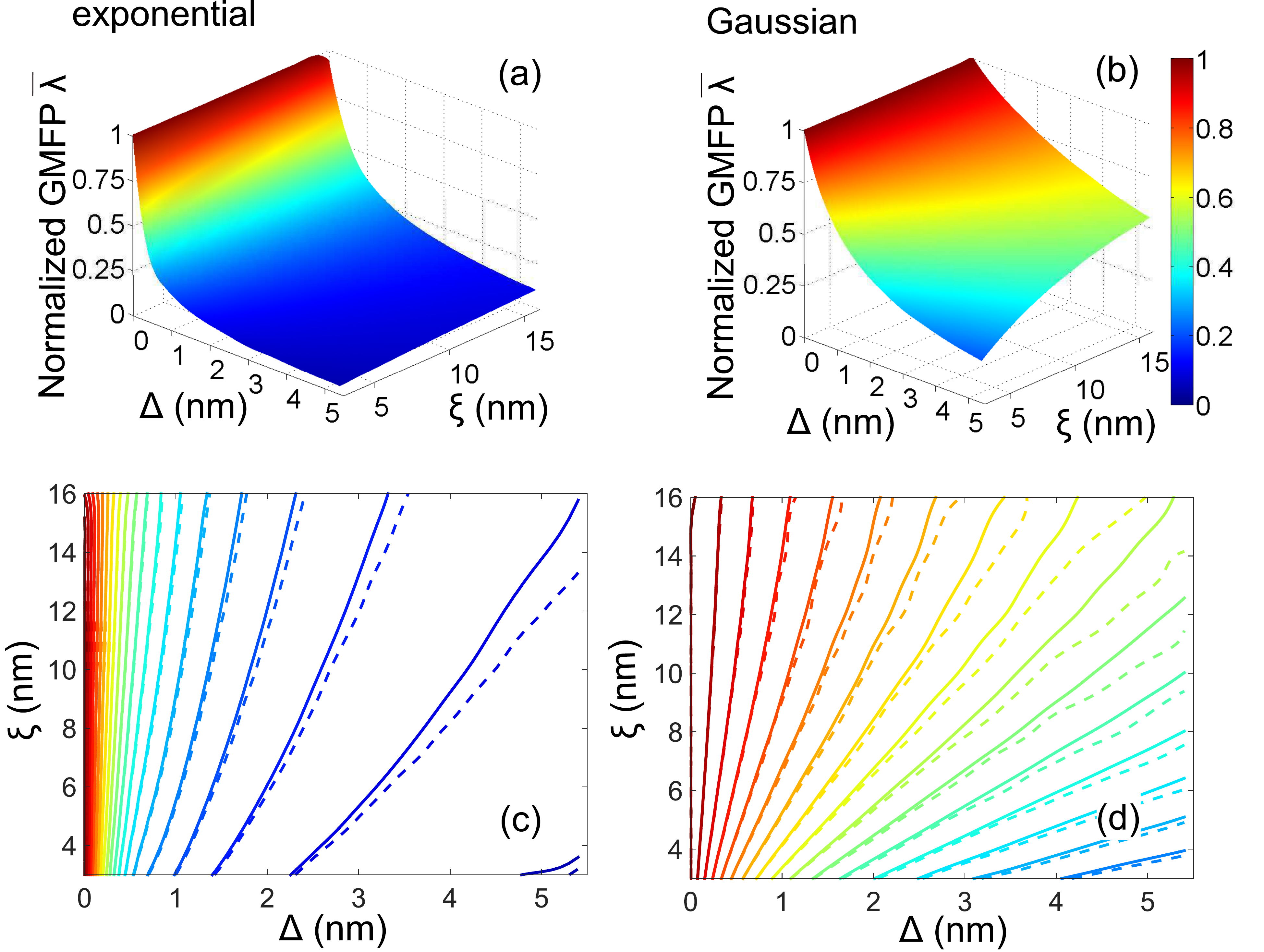}
\caption{(Top row) $\bar\lambda$, the geometric mean free path normalized to the SiNW width, as a function of rms roughness $\Delta$ and correlation length $\xi$ for a 70-nm-wide SiNW whose rough boundary surfaces are characterized by (a) exponential and (b) Gaussian autocorrelation functions. (Bottom row) Contour plots of the normalized geometric mean free path $\bar\lambda$ versus $\Delta$ and $\xi$ for exponential (c) and Gaussian (d) boundary surfaces. Here, solid lines correspond to NWs of width 70 nm, while dashed lines represent 35-nm-wide NWs. Consecutively colored contours correspond to 0.05 increments in $\bar\lambda$. The color scale is the same as in the top row.}\label{fig:GMFP-vs-delta-and-l}
\end{figure}

Figure \ref{fig:GMFP-vs-delta-and-l} shows $\bar\lambda $  as a function of
$\Delta$ and $\xi$ for exponential and Gaussian correlation on 70-nm-wide SiNWs. Note that the $\bar{\lambda}(\Delta,\xi)$ surfaces look qualitatively similar to $\kappa (\Delta,\xi)$ (Fig. \ref{fig:Kappa-vs-rms-and-clen}); both $\kappa$ and $\overline{\lambda}$
decrease as $\Delta$ increases and increase as $\xi$
increases. The bottom two panels in Fig. \ref{fig:GMFP-vs-delta-and-l} reveal an important universality: the contour plots of $\bar{\lambda}(\Delta,\xi)$ for 35 and 70-nm-wide NWs are nearly identical for a given ACF. Therefore, once we specify the type of correlation, the normalized geometric mean free path $\bar{\lambda}$ can be considered as a near-universal quantifier of surface roughness scattering, encompassing $W$, $\Delta$, and $\xi$. From this point on, we will concentrate on exponential surfaces, as they resemble experiment more closely. \cite{PYang_NLet_12}

\begin{figure}
\includegraphics[width=0.8\columnwidth]{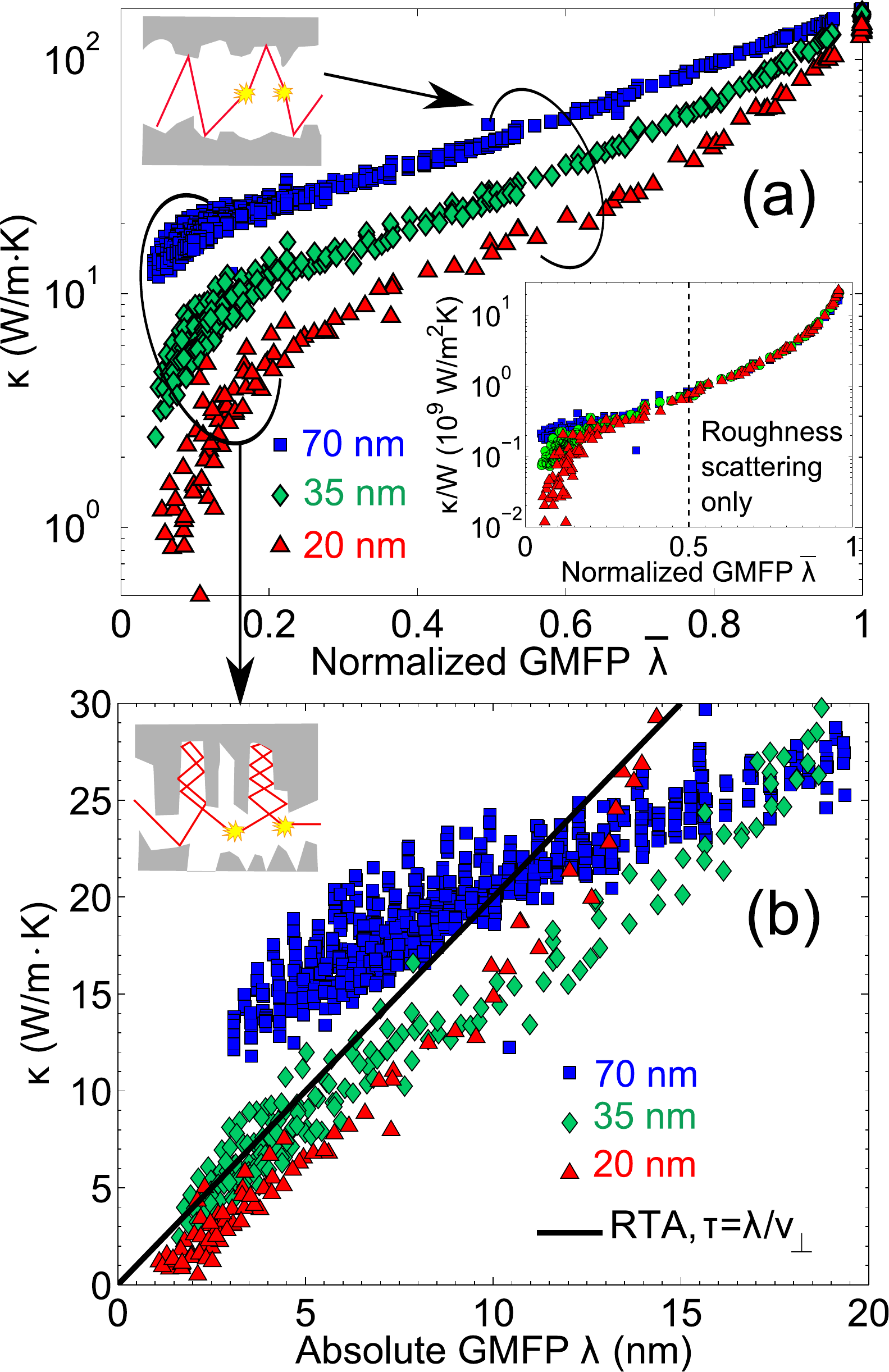}
\caption{(a) Room-temperature thermal conductivity $\kappa$ as a function the normalized geometric mean free path $\bar\lambda$ for SiNWs of width 70 nm (blue squares), 35 nm (green diamonds), and 20 nm (red triangles), obtained from PMC simulation on a large ensemble of SiNWs with real-space exponentially correlated rough surfaces. (Inset) $\kappa$ normalized to width $W$ versus $\bar\lambda$ without internal scattering for the same NWs as in the main panel. Vertical dashed line denotes $\bar\lambda=1/2$. (b) Thermal conductivity $\kappa$ as a function of the absolute geometric mean free path $\lambda$ for very rough SiNWs [low-$\bar\lambda$ region from (a)]. The solid black line corresponds to a simple RTA solution with $\tau=\lambda/v_{\perp}$  that yields a thermal conductivity of $\kappa = A\lambda$, $A \approx 2\times 10^9\,\mathrm{W/m^2K}$. The schematics in the top left corners depict phonon trajectories, interrupted by internal scattering events, in NWs with (a) low-to-moderate and (b) pronounced roughness.}\label{fig:kappa-vs-gmfp}
\end{figure}

Figure \ref{fig:kappa-vs-gmfp}(a) shows $\kappa$ as a function of
$\overline{\lambda}$ for SiNWs of width 20, 35, and 70 nm and with exponential surfaces. Each plot presents thermal conductivity data for several hundred different NWs with real-space roughness. There are a number of interesting features in this graph. First, as $\bar{\lambda}\rightarrow 1$ (the smooth-surface limit), all three curves converge to the bulk thermal conductivity, \cite{Slack_PR_64} as expected. This smooth-wire limit is well captured by treating each surface with an appropriate specularity parameter, as seen in Fig. \ref{fig:kappa-Soffer}. Second, the Casimir limit values (17.3, 26.1, and 42.7 $\mathrm{W/m\cdot K}$ for the 20, 35, and 70-nm NWs, respectively), obtained from PMC simulations with smooth but momentum-randomizing surfaces, occurs at $\bar{\lambda}\approx 1/2$, i.e. $\lambda=W/2$. Therefore, thermal conductivity below the Casimir limit corresponds to $\bar\lambda<1/2$.

In the inset to Fig. \ref{fig:kappa-vs-gmfp}(a) we present thermal conductivity normalized to width, $\kappa/W$, versus $\bar\lambda$ when internal scattering mechanisms have be turned off and only surface-roughness scattering remains. For $\bar\lambda>1/2$, i.e. above the Casimir limit, the curves for the three different NW widths coincide: roughness-limited $\kappa/W$ is universal as a function of $\bar\lambda$ in this regime. With increasing roughening, however, once the Casimir limit has been surpassed ($\bar\lambda<1/2$), the curves start to diverge from one another. Therefore, there is a fundamental difference in surface-limited transport above and below the Casimir limit, an important issue that we explore further below.

The most prominent feature of Fig. \ref{fig:kappa-vs-gmfp}(a) is that, for each NW thickness, there is a crossover in $\kappa(\bar{\lambda})$. In the low-to-moderate roughness regime, $\kappa (\bar\lambda)$ is exponential ($\ln\kappa\sim \bar\lambda$, with a width-dependent slope), notably so for the thicker two NWs (35 and 70 nm). Considering that these are the results of semiclassical simulations, localization \cite{Anderson1958} stemming from coherent superposition of waves is out of the question. We believe the explanation is related to the phenomenon of variable-range hopping, \cite{Mott69,Apsley74} where there is an exponential dependence of conductivity of the characteristic hopping range. What happens here, for low-to-moderate roughness,  can be referred to as \textit{variable-range bouncing}: phonon bounces between opposite sides of the NW, while the range variability stems from the path being interrupted by internal scattering events, as depicted in the top-left-corner schematic in Fig. \ref{fig:kappa-vs-gmfp}(a).

Another interesting aspect of Fig. \ref{fig:kappa-vs-gmfp}(a) is the $\kappa(\bar{\lambda})$ behavior of very rough NWs (small $\bar\lambda$). First off, the cross-over happens at smaller $\bar\lambda$ for thicker NWs, because in thicker NWs the relative importance of internal scattering at a given $\bar\lambda$ is higher. The small-$\bar\lambda$ region is also harder to reach in thicker NWs, like the 70-nm one, because very high $\Delta$ would be required, outside of the range we focused on here. The $\kappa(\bar{\lambda})$ behavior in the small-$\bar\lambda$ region can be explained by \textit{multiple scattering events at the same boundary}. In Fig. \ref{fig:kappa-vs-gmfp}(b), we zoom in on the low-$\bar\lambda$ region from Fig. \ref{fig:kappa-vs-gmfp}(a) and present $\kappa$ as a function of the absolute GMFP, $\lambda$. We note that the thermal conductivities reach very low values, of only a few W/m$\cdot$K, such as those measured by Hochbaum \textit{et al}. \cite{PYang_Nat_08} We also see that the data for 20-nm and 35-nm NWs fall on top of each other, which is in keeping with the intuitive picture that multiple scattering events from the same boundary govern transport, so crossing the wire, and thus the wire width, becomes irrelevant. Indeed, the low-$\lambda$ region agrees very well with the simple relaxation-time approximation (RTA) expression $\kappa=(2\pi)^{-3}\sum_{b}\int d^3\vec{q}\,c(b,\vec{q})v^2_{||}(b,\vec{q})\tau (b,\vec{q})$, where we put $\tau (b,\vec{q})=\lambda/v_{\perp}(b,\vec{q})$. Here, $c(b,\vec{q})$ is the heat capacity of mode $b$ at wave number $\vec{q}$, while $v_{||}(b,\vec{q})$ and $v_{\perp}(b,\vec{q})$ are the velocity components along and across the wire, respectively. Upon simple integration, we obtain  $\kappa=A\lambda$, where $A\approx 2\times 10^9\,\mathrm{W/m^2 K}$ at 300 K [Fig. \ref{fig:kappa-vs-gmfp}(b)]. This $\lambda$ is single-surface dominated [see top-left-corner schematic in Fig. \ref{fig:kappa-vs-gmfp}(b)] and bears essentially no dependence on the NW width. We note that Sadhu and Sinha \cite{Sinha_PRB_11} argued that coherent transport and multiple correlated scattering events at a surface are key to low thermal conductivities. Here, we obtain ultralow thermal conductivities within a semiclassical (incoherent) transport model. While coherent transport is clearly not required for ultralow $\kappa$, multiple scattering processes from a surface are.



In summary, we showed that thermal conductivity $\kappa$ far below the Casimir limit can be readily obtained for NWs within the Boltzmann transport formalism,  provided that real-space rough surfaces with realistic ACFs are employed. We introduced the concept of the normalized geometric mean free path $\bar\lambda$, which encompasses NW width, roughness rms height, and correlation length. Thermal conductivities below the Casimir limit correspond to $\bar\lambda<1/2$. $\kappa(\bar{\lambda})$ reveals universal signatures of the interplay between boundary-roughness and internal scattering: a) in the low-roughness, high-$\bar\lambda$ region, $\kappa(\bar{\lambda})$ is exponential, owing to  phonons bouncing across the NW and having trajectories randomly interrupted by internal scattering events; b) in the high-roughness, low-$\bar\lambda$ region, multiple scattering events at a single interface govern transport, $\kappa\sim\bar{\lambda}$, and extremely low values of $\kappa$, close to the amorphous limit, are obtained. This work shows that pronounced roughness results in a fundamental, qualitative change to thermal transport in nanostructures.

Finally, while the NWs with our numerically generated surfaces are technically not truly chaotic cavities, \cite{Chernov_Book,VegaPRE93,TroubetzkoyChaos98,Alonso_PRE_2002} the real NWs likely are. \cite{PYang_NLet_12} An individual phonon in a real SiNW may exhibit classically chaotic \cite{Alonso_PRL_1999} or quantum-chaotic dynamical features, \cite{Tanner_JoPA_2007} depending on its wave length and whether coherence is preserved upon surface scattering. \cite{CahillAPR2012} A deeper understanding of thermal transport in very rough NWs means understanding how ensembles of phonons behave inside chaotic cavities. \cite{CecconiChaos2005}

The authors gratefully acknowledge support by the U.S. Department of Energy, Office of Science, Materials Science Program, award DE-SC0008712. The work was performed using the resources of the UW-Madison Center For High Throughput Computing (CHTC).

%

\end{document}